\documentclass[prd,aps,10pt,twocolumn,showpacs,preprintnumbers,superscriptaddress]{revtex4-1}

\usepackage{amsmath,amssymb,amsfonts,amsthm}
\usepackage{amsbsy} 
\usepackage{epsfig}
\usepackage{latexsym}
\input amssym.def
\input amssym.tex

\usepackage{color}
\usepackage{hyperref}

\def\com{\color{magenta}}
\def\cob{\color{blue}}
\newcommand{\oarX}[1]{\href{http://arxiv.org/abs/#1}{{\ttfamily\com #1}}}
\newcommand{\arX}[1]{\href{http://arxiv.org/abs/#1}{{\ttfamily\com arXiv:#1}}}
\newcommand{\doin}[2]{\href{http://dx.doi.org/#1}{\cob #2}}

\def\barr{\begin{array}}
\def\earr{\end{array}}
\def\half{\frac{1}{2}}
\def\ben{\begin{equation}}
\def\een{\end{equation}}
\def\bs{\begin{subequations}}
\def\es{\end{subequations}}
\def\bena{\begin{eqnarray}}
\def\eena{\end{eqnarray}}

\def\bC{\mathbb{C}}

\def\SU{{\rm SU}}

\def\im{{\rm i}}

\newcommand{\dd}{\mathrm{d}}
\newcommand{\hphi}{\hat{\varphi}}
\newcommand{\hphid}{\hat{\varphi}^{\dagger}}

\begin{document}
\title{Perturbing a quantum gravity condensate}
\author{\bf Steffen Gielen}
\affiliation{Theoretical Physics, Blackett Laboratory, Imperial College London, London SW7 2AZ, United Kingdom}
\email{s.gielen@imperial.ac.uk}
\affiliation{Perimeter Institute for Theoretical Physics, 31 Caroline Street North, Waterloo, Ontario N2L 2Y5, Canada} 

\begin{abstract}

In a recent proposal using the group field theory approach, a spatially homogeneous (generally anisotropic) universe is described as a quantum gravity condensate of ``atoms of space,'' which allows the derivation of an effective cosmological Friedmann equation from the microscopic quantum gravity dynamics. Here we take a first step towards the study of cosmological perturbations over the homogeneous background. We consider a state in which a single ``atom'' is added to an otherwise homogeneous condensate. Backreaction of the perturbation on the background is negligible and the background dynamics can be solved separately. The dynamics for the perturbation takes the form of a quantum cosmology Hamiltonian for a ``wave function,'' depending on background and perturbations, of the product form usually assumed in a Born--Oppenheimer approximation. We show that the perturbation we consider corresponds to a spatially homogeneous metric perturbation, and for this case derive the usual procedures in quantum cosmology from fundamental quantum gravity.
\end{abstract}

\date{3 February 2015}

\preprint{Imperial/TP/2014/SG/1}

\pacs{98.80.Qc, 04.60.Pp, 98.80.Bp}

\maketitle

\section{Introduction}

The most natural point of contact between observable phenomena and fundamental theories of quantum gravity is probably in the cosmology of the early Universe. In spite of the phenomenological successes of inflation as a theory of the early Universe, classically a generic inflationary universe must have emerged from a singularity, implying a breakdown of classical general relativity \cite{singu}. Quantum gravity could also provide insight in the search for a theory of initial conditions for the Universe. In practice, describing cosmological singularities and, more generally, cosmologically interesting time-dependent spacetimes has been a difficult task in basically all approaches to quantum gravity. In loop quantum gravity (LQG) \cite{LQG} the task is complicated by the property of background independence which implies that the natural (Ashtekar--Lewandowski) vacuum \cite{ashlewa} describes a completely degenerate (metric) geometry; a state describing a macroscopic and approximately smooth geometry cannot be found as a small perturbation of this vacuum state. The Dittrich--Geiller vacuum \cite{dgvac}, describing a flat connection but completely undetermined metric, seems a more promising starting point, but the restriction to exactly flat geometries is from the perspective of cosmology rather severe.

In the absence of a fully satisfactory description of cosmological spacetimes within quantum gravity, a common strategy is to perform a symmetry reduction at the classical level and to quantize only the degrees of freedom of the reduced system. This leads to minisuperspace models of Wheeler--DeWitt quantum cosmology, with a long history \cite{qcrev}, or, when LQG techniques are used in the quantization, to loop quantum cosmology (LQC) \cite{LQCrev}. LQC models confirm the expectation that the classical singularity is resolved by quantum gravity effects, leading to a big bounce. Their precise relation to the full theory of LQG has however not been fully clarified so far.

A new proposal addressing this fundamental issue was put forward in Refs.~\cite{gfcpapers}. Working in the group field theory (GFT) approach to quantum gravity, itself a second quantization formulation for LQG \cite{daniele2q} (such a second quantization of what is already a field theory is sometimes called ``third quantization'' \cite{3rdquan}), the new idea is the description of a spatially homogeneous (generally anisotropic) universe as a {\em condensate} of elementary excitations of quantum geometry, or ``atoms of space.'' As the number of atoms in such a condensate is taken to be very large, it can be interpreted as an approximate continuum spacetime; the property of condensation, implying that all microscopic geometric degrees of freedom are in the same quantum state, is analogous to spatial homogeneity for a continuum manifold. 

The results of Refs.~\cite{gfcpapers} show that the description of space as a quantum gravity condensate goes beyond a purely kinematical construction. Imposing some of the GFT Schwinger--Dyson equations as conditions on a given condensate state, and hence demanding that the condensate is, for the operators chosen, a good approximation to a nonperturbative GFT vacuum, leads to conditions on the ``condensate wave function'' that can be interpreted as (generally nonlinear, nonlocal) effective quantum cosmology equations where the condensate wave function plays the role of a quantum cosmology wave function. (A different approach, using the notion of {\em fidelity} instead of Schwinger--Dyson equations for deriving an effective dynamics, is outlined in Ref.~\cite{lorenzofid}.)

The resulting effective dynamical equations for GFT condensate states are still fully quantum. In Refs.~\cite{gfcpapers}, a semiclassical WKB limit was used to interpret them in terms of classical cosmological dynamics, and it was shown that, in the isotropic case, they reduce exactly to the Friedmann equation of general relativity. This result was obtained both in Riemannian and Lorentzian signature, for pure vacuum and for gravity with a massless scalar field. The meaning of the WKB approximation in this context has however been debated. Exact solutions for isotropic universes (even if they are oscillatory) can deviate strongly from the WKB expectations \cite{steffegfc}; if the scaling of macroscopic observables with the number of atoms in the condensate is taken into account, the WKB expansion in derivatives appears to be an expansion in the ratio of the Planck area to the {\em average} area per atom, which is not necessarily a small parameter \cite{atomnumber}. Accounting for this scaling, an effective cosmological dynamics can be derived from expectation values of macroscopic (cosmological) observables without any semiclassical approximation; the interpretation of the resulting equations in terms of cosmological variables such as the scale factor then depends on how the number of atoms scales with the cosmological variables, and is not necessarily compatible with the WKB results \cite{atomnumber}. See however also Ref.~\cite{gianluca} for how the dynamics of isotropic LQC can emerge from a WKB limit of effective equations of GFT condensates.

While the use of quantum gravity condensates offers promising opportunities for deriving quantum cosmology models from a more fundamental theory, the formalism so far suffers from a basic restriction, as the assumed condensate states correspond to an exactly spatially homogeneous universe. In the geometric interpretation of general many-atom GFT states given in Refs.~\cite{gfcpapers}, the geometric data contained in such a state specifies a metric on a continuum manifold after an embedding of the basic geometric quanta (interpreted as elementary tetrahedra, or simplicial manifolds composed of a few tetrahedra) into a given 3-manifold is chosen. For a condensate, where all basic quanta are in the same state, the choice of embedding is arbitrary, as is consistent with invariance of the metric under a transitive group action; for a more general geometry, however, the reconstructed metric depends on the choice of embedding. It is not yet clear how to relax the assumption of strict homogeneity, even perturbatively, as one would need to do to incorporate cosmological perturbations into GFT many-atom states, and to connect with the usual formalism \cite{cosmpert}.

In quantum cosmology, one standard procedure for including inhomogeneities is to follow Ref.~\cite{hallhawk}. One considers a perturbed homogeneous, isotropic Friedmann--Lema\^itre--Robertson--Walker (FLRW) universe, expands the Hamiltonian up to second order in the perturbations, and, as the different fluctuation modes are not coupled to each other, assumes a wave function of product form
\ben
\Psi(a,\phi,x_n)=\Psi_0(a,\phi)\prod_n \Psi^{(n)}(a,\phi;x_n)
\label{prodform}
\een
where $a$ and $\phi$ are the scale factor and scalar field of the background and $n$ labels the fluctuation modes. One then works in a Born--Oppenheimer approximation where the Wheeler--DeWitt equation for the background wavefunction $\Psi_0$ is solved separately from the fluctuations which propagate on a semiclassical background, given by $\Psi_0$ in a WKB limit. For a recent application of this formalism to a computation of quantum gravitational corrections to the cosmic microwave background (CMB) spectrum, see e.g. \cite{kiefkr}. In such calculations in quantum cosmology, a number of assumptions have to be made regarding the smallness of fluctuations with respect to the background and the applicability of the Born--Oppenheimer approximation. There is no embedding of wave functions into a well-defined Hilbert space in which the error of these approximations could be quantified; this provides one of the main motivations for seeking to derive quantum cosmology models consistently from some candidate theory of quantum gravity.

In this paper, we take a first step towards extending the proposal of quantum gravity condensates to cosmological perturbations. Using the formalism of group field theory and its Fock space of atoms of space, we consider the simplest possible perturbation of a fully homogeneous condensate, a state in which an elementary excitation is added to the condensate. This state is characterized by two separate wave functions for the condensate and the perturbation. We compute its effective dynamics by using Schwinger--Dyson equations of the GFT as proposed in Refs.~\cite{gfcpapers, atomnumber}, and find confirmation of several assumptions made in the same context in quantum cosmology: First, the backreaction of the perturbation on the condensate is negligible, so that the dynamics for the ``background'' condensate wave function can be solved separately. Then the resulting dynamics for the perturbation takes the form of a Wheeler--DeWitt Hamiltonian for a wave function which is the product of the wave functions for condensate and perturbations. The Hamiltonian for this product wave function contains one part for the background plus a part for the perturbations of identical form. We explain why the interpretation of such a perturbation as spatially homogeneous is fully consistent with the geometric interpretation given in Refs.~\cite{gfcpapers}. We stress that none of these results arise from the assumption that our perturbed condensate describes a perturbed FLRW universe; instead they are {\em derived} from the kinematics and dynamics of a full theory of quantum geometry. They provide reassuring consistency with conventional quantum cosmology.

While the perturbation we consider can classically be absorbed into the background, quantum mechanically background and perturbation can be distinguished by the different number of quanta of geometry. Our example, while restricted to a special case, exemplifies the possibility for adding perturbations to quantum gravity condensates, to be investigated further in future work.

\section{Quantum gravity condensates and Cosmology}
\label{condenintro}

In this section we give a short summary of the proposal of Refs.~\cite{gfcpapers} for describing spatially homogeneous universes by condensate states in group field theory.

The GFT formalism itself \cite{gftreviews} was developed as a covariant quantum field theory formulation of the dynamics of loop quantum gravity. In LQG, transition amplitudes for boundary spin network states, interpreted as discrete geometries, are given in terms of a {\em spin foam amplitude} associated to each discrete spacetime history that interpolates between the prescribed boundary data \cite{LQG}. In GFT, the same amplitudes are generated as Feynman amplitudes associated to the discrete spacetime histories appearing as Feynman graphs. Spin foam models and GFT actions are in one-to-one correspondence \cite{gftspinf}.

Just as in condensed matter physics, using a second quantized quantum field theory formulation of the dynamics of LQG gives access to a variety of techniques and simplifies many considerations. In particular, in analogy to the physics of Bose--Einstein condensates, one can define a {\em condensate} of atoms of geometry. In this picture, a homogeneous universe is made up of many disconnected discrete geometric building blocks, all in the same microscopic quantum state, so that they carry the same geometric information. The (approximate) metric one reconstructs from such a discrete geometry is spatially homogeneous \cite{gfcpapers}. As in Bose--Einstein condensates, the GFT field operator acquires a nonzero expectation value which is interpreted as a quantum cosmology wave function, subject to nonlinear equations of motion analogous to the Gross--Pitaevskii equation. Such equations can then be interpreted in terms of cosmological observables, for instance by considering expectation values or a semiclassical approximation.

More concretely, the kinematical Hilbert space of discrete geometries in GFT can be defined as a Fock space. One starts with a Fock vacuum $|\emptyset\rangle$ which is analogous to the Ashtekar--Lewandowski vacuum of LQG; it corresponds to a completely degenerate geometry, with zero expectation value for all areas and volumes, and no excitations of quantum geometry. (That such a state gives a natural vacuum can be understood by observing that only a zero metric is invariant under diffeomorphisms.) There is a basis of creation operators that create excitations when acting on $|\emptyset\rangle$. In three spatial dimensions, these excitations are interpreted as tetrahedra with geometric information attached to them. In the ``group'' representation, four group elements $g_I$ define the parallel transports of a gravitational connection along links dual to the four faces; in the dual ``metric'' representation the data is given by four Lie algebra elements $B_I$ corresponding to the area element integrated over the four faces, $B^{AB}_I\sim\int_{\triangle_I} e^A\wedge e^B$.

One particular set of one-particle states is given by acting with the GFT field operator, in the group representation, on $|\emptyset\rangle$,
\ben
|g_1,\ldots,g_4\rangle:=\hat\varphi^{\dagger}(g_1,\ldots,g_4)|\emptyset\rangle\,.
\een
Such a state is interpreted as a single tetrahedron with discrete geometric data given by the group elements $g_I$ interpreted as parallel transports of a connection. Consequently, the $g_I$ take values in a group $G$ interpreted as the gauge group of gravity. Depending on the model, one usually takes $G={\rm Spin}(4)$, $G={\rm SL}(2,\bC)$ or $G=\SU(2)$, which is the gauge group in the Ashtekar--Barbero formulation of gravity, and hence in LQG.

The Fock space can now be constructed by repeated actions of the field $\hat\varphi^{\dagger}(g_I)$, taking into account the (nonrelativistic) bosonic commutation relations
\bena
\left[\hat\varphi(g_I),\hat\varphi(g'_I)\right]=\left[\hat\varphi^\dagger(g_I),\hat\varphi^\dagger(g'_I)\right] & = & 0\,,\nonumber
\\ \left[ \hat{\varphi}(g_I),\hat{\varphi}^{\dagger}(g'_I)\right] & = & {\bf 1}_G (g_I,g'_I)\,,
\label{commrel}
\eena
where ${\bf 1}_G$ is a gauge-invariant delta distribution. In the rest of the paper, we will assume compact $G$, with a normalized Haar measure $\int {\rm d}g = 1$. We can then set ${\bf 1}_G(g_I,g'_I):=\int {\rm d}h\;\delta^4(g_Ih{g'_I}^{-1})$. Equation (\ref{commrel}) is compatible with the gauge invariance property of the field $\hat\varphi$,
\ben
\hat\varphi(g_1,\ldots,g_4)=\hat\varphi(g_1h,\ldots,g_4h)\quad\forall\;h\in G\,,
\label{gaugeinv}
\een
which corresponds to invariance of the theory under gauge transformations acting on a vertex where all four links associated to a tetrahedron meet; these act as $g_I\mapsto g_I\,h$ as in lattice gauge theory.

Any given $N$-particle state in the Fock space is interpreted as a geometric structure made up of $N$ tetrahedra with discrete geometric data. Depending on the state, these can be connected, with several or all faces glued to one another, or disconnected. In any case, {\em a priori} they are not embedded in any ``space,'' but themselves make up space and its geometry. GFTs are not quantum field theories {\em on} space but {\em of} space. The domain space of the field $\hat\varphi$ is the abstract group manifold $G^4$ which is the configuration space of a single tetrahedron, and has no relation to space or spacetime.

In Refs.~\cite{gfcpapers}, an embedding into a given manifold (of fixed topology) was used in order to reconstruct an approximate metric geometry from given GFT Fock states. In general, the reconstructed metric depends on the choice of embedding, which is arbitrary. However, for a spatially homogeneous metric it does not, as a homogeneous geometry can be fully reconstructed from any one given point. The criterion for spatial homogeneity is that all GFT quanta carry the same geometric data, which is analogous to the condition of condensation in condensed matter systems. By taking the (average) particle number $N$ as large as possible, the approximate metric reconstructed from the discrete GFT data gives an arbitrarily good approximation to a continuum metric.

There are a few ambiguities in the definition of GFT condensate states. In particular, one can consider a condensate of ``atoms,'' single tetrahedra, or a condensate of ``molecules'' which are composed of two or more tetrahedra. The simplest type of molecule would be a ``dipole'' of two tetrahedra with all four faces pairwise identified, which is the simplest triangulation of a three-sphere. The dipole is the elementary building block in the spin foam cosmology approach \cite{sfcosmo}, which also aims at describing spatially homogeneous universes within LQG. In Refs.~\cite{gfcpapers}, both types of condensates were considered. Condensates of atoms are much simpler to handle technically, as the associated quantum states are coherent states of the GFT field operator. For the exploratory purposes of this paper, we will only consider this type of condensate.

The unperturbed condensate state is then defined by
\ben
|\sigma\rangle := \mathcal{N}(\sigma)\exp\left(\hat\sigma\right)|\emptyset\rangle\,,
\label{conddef}
\een
where 
\ben
\hat\sigma:=\int (\dd g)^4\;\sigma(g_I)\;\hphid(g_I)
\een
and $\mathcal{N}(\sigma)$ is a normalization factor, and without loss of generality $\sigma(g_I)=\sigma(g_Ih)$ for all $h\in G$, due to Eq.~(\ref{gaugeinv}). $\mathcal{N}(\sigma)$ can be computed by noting that
\ben
\langle\emptyset|\exp\left(\hat\sigma^\dagger\right)\exp\left(\hat\sigma\right)|\emptyset\rangle = \exp\left(\int (\dd g)^4\;|\sigma(g_I)|^2\right)
\een
and hence 
\ben
\mathcal{N}(\sigma):=\exp\left(-\half\int (\dd g)^4\;|\sigma(g_I)|^2\right)\,.
\een
It is then immediate to verify that $|\sigma\rangle$ indeed satisfies 
\ben
\hat\varphi(g_I)|\sigma\rangle = \sigma(g_I)|\sigma\rangle\,.
\een
Using this, the average particle number is
\ben
N:=\int (\dd g)^4\;\langle\sigma|\hphid(g_I)\hphi(g_I)|\sigma\rangle =  \int (\dd g)^4\;|\sigma(g_I)|^2\,.
\een
Hence the integral of $\sigma$ is {\em not} normalized to one, but corresponds to a physical observable of the condensate. As mentioned above, in order for the discrete spatial geometry formed by the condensate to be a good approximation to a continuum homogeneous universe one needs $N\gg 1$ (e.g. $N$ could be the volume of the spatial region of interest in Planck units). There may be constraints on the possible values for $N$ coming from the dynamics of the given GFT model, through the requirement for (\ref{conddef}) to be a good approximation to a physical state.

This requirement can be expressed a set of equations for the condensate wave function $\sigma(g_I)$ which can be derived, among other means, from Schwinger--Dyson equations of the GFT. These equations can be formally derived from the path integral, and require expectation values of certain operators to vanish in any vacuum state of the theory (itself defined through the path integral). 

The simplest such operator is the equation of motion
\ben
\left\langle\frac{\delta S[\varphi,\bar\varphi]}{\delta\bar\varphi(g_I)}\right\rangle = 0\,.
\label{eqofmo}
\een
More generally, one can insert an operator $\mathcal{O}[\varphi,\bar\varphi]$ into the path integral to find additional relations of the form
\ben
\left\langle\frac{\delta \mathcal{O}[\varphi,\bar\varphi]}{\delta\bar\varphi(g_I)}-\mathcal{O}[\varphi,\bar\varphi]\frac{\delta S[\varphi,\bar\varphi]}{\delta\bar\varphi(g_I)}\right\rangle = 0\,.
\label{schwdys}
\een
The idea is now to use expectation values such as Eq.~(\ref{schwdys}), evaluated in the state $|\sigma\rangle$, as information about the underlying GFT dynamics, and to interpret the resulting equations for the condensate wave function $\sigma$ as quantum cosmology equations. The simplest such equation Eq.~(\ref{eqofmo}) would, in the case of a Bose--Einstein condensate, precisely reproduce the Gross--Pitaevskii equation for the condensate wave function $\Psi$.

Here we follow the approach introduced in Ref.~\cite{atomnumber} and interpret (\ref{schwdys}) as the expectation value of a suitable many-body operator on the GFT Fock space. One sets $\mathcal{O}=\bar\varphi(g_I)$ and integrates over the $g_I$. Under normal ordering, the delta distribution $\delta\bar\varphi/\delta\bar\varphi$ disappears and one obtains, in terms of normal ordered operators,
\ben
\langle\hat{K}\rangle + \left\langle \int ({\rm d}g)^4\;\hat\varphi^{\dagger}(g_I)\frac{\delta\hat{\mathcal{V}}[\hat\varphi,\hat\varphi^\dagger]}{\delta\hat\varphi^{\dagger}(g_I)}\right\rangle = 0 
\label{expvalue}
\een
where we have written the GFT action as $S=K+\mathcal{V}$ with quadratic kinetic term $K$ and potential $\mathcal{V}$. As in Ref.~\cite{atomnumber}, we now also assume that the second term in Eq.~(\ref{expvalue}) vanishes. This can be an exact result for a certain class of states, such as the dipole condensate states defined in Refs.~\cite{gfcpapers}, or more generally correspond to a weak-coupling limit in which the GFT interactions are neglected. Neglecting the second term in Eq.~(\ref{expvalue}) also simplifies the quantum cosmology interpretation, as Eq.~(\ref{eqofmo}) becomes a linear equation of motion for $\sigma$, as in standard quantum cosmology. (The more general, nonlinear case could be related to the nonlinear extension of quantum cosmology introduced in Ref.~\cite{nonl}.).

The expectation value of the GFT ``kinetic energy'' is
\bena
\langle\hat{K}\rangle&:=&\int (\dd g)^4\;\langle\sigma|\hphid(g_I)\mathcal{K}\hphi(g_I)|\sigma\rangle \nonumber
\\&=&  \int (\dd g)^4\;\bar\sigma(g_I)\mathcal{K}\sigma(g_I)\,,
\eena
where we are assuming a local kinetic term specified by the choice of a differential operator $\mathcal{K}$ on $G^4$. 

Imposing the requirement $\langle\hat{K}\rangle = 0$ on the condensate trial state (\ref{conddef}) can thus be interpreted as a given many-body operator on the GFT Fock space having zero expectation value. The classical limit of this operator can be interpreted as an effective Hamiltonian constraint corresponding to a generalized Friedmann equation, written in terms of cosmological observables such as the scale factor and Hubble ``parameter,'' as detailed in Ref.~\cite{atomnumber}. This provides the link between the Schwinger--Dyson equations of the GFT and an effective quantum cosmology equation, and hence between the microscopic dynamics of quantum geometry and large-scale cosmological dynamics, in a way analogous to deriving an effective hydrodyamic description (e.g.~the Euler equation) of a quantum fluid by using coherent states in condensed matter physics. See Refs.~\cite{gfcpapers} for more details and conceptual background.

\section{Adding a perturbation}

GFT condensate states such as Eq.~(\ref{conddef}) can be interpreted as spatially homogeneous geometries. Although an embedding of the quanta of geometry into a manifold is used in the reconstruction of an approximately smooth geometry defined by the quantum state, the property of condensation, meaning that all quanta are in the same microscopic state, makes the reconstructed geometry independent of the embedding. This convenient feature of the exact condensate is however rather restrictive; our Universe is not exactly homogeneous, and being able to reproduce the correct spectrum of cosmological perturbations is an important consistency check for any proposed model of quantum cosmology.

Developing a formalism for the study of cosmological perturbations over exactly homogeneous condensates in quantum gravity will require new conceptual insights. There is no obvious notion of coordinates for the condensate with respect to which perturbations could be localized; on the contrary, the condensate is made up of indistinguishable quantum particles. One expects an effective classical picture of a background (e.g.~FLRW) geometry to be meaningful only for a condensate with semiclassical properties; appropriate conditions for semiclassicality must presumably be defined for macroscopic instead of microscopic observables, as there is no reason to expect semiclassical behavior at the Planck scale (as discussed in Refs.~\cite{steffegfc,atomnumber}).

In absence of a complete picture, we will take a first step into the study of perturbations of homogeneous GFT condensates by considering the simplest possible type of perturbation. Namely, we take the state (\ref{conddef}) and create another elementary excitation over it,
\ben
|\tau,\sigma_0\rangle := \mathcal{N}(\tau,\sigma_0)\,\hat\tau\,\exp\left(\hat\sigma_0\right)|\emptyset\rangle\,,
\label{pertcond}
\een
where we define
\ben
\hat\tau:=\int (\dd g)^4\;\tau(g_I)\;\hphid(g_I)\,,
\een
we use the notation $\sigma_0$ instead of $\sigma$ to emphasize this specifies the background, and $\tau$ is the wave function for the additional excitation. Computing the normalization factor $\mathcal{N}(\tau,\sigma_0)$, we find
\bena
&&\langle\emptyset|\exp\left(\hat\sigma_0^\dagger\right)\hat\tau^\dagger\hat\tau\exp\left(\hat\sigma_0\right)|\emptyset\rangle\nonumber
\\&=&\langle\emptyset|\exp\left(\hat\sigma_0^\dagger\right)\hat\tau\hat\tau^\dagger\exp\left(\hat\sigma_0\right)|\emptyset\rangle\nonumber
\\&&+\int (\dd g)^4 \;|\tau(g_I)|^2\,\langle\emptyset|\exp\left(\hat\sigma_0^\dagger\right)\exp\left(\hat\sigma_0\right)|\emptyset\rangle\nonumber
\\&=&\left(\left|\int (\dd g)^4\;\bar\tau(g_I)\sigma_0(g_I)\right|^2+\int (\dd g)^4 \;|\tau(g_I)|^2\right)\nonumber
\\&&\times\exp\left(\int (\dd g)^4\;|\sigma_0(g_I)|^2\right)\,,
\eena
and so
\bena
\mathcal{N}(\tau,\sigma_0)&=&\left(\left|\int (\dd g)^4\;\bar\tau(g_I)\sigma_0(g_I)\right|^2+\int (\dd g)^4 \;|\tau(g_I)|^2\right)^{-\frac{1}{2}}\nonumber
\\&&\times\exp\left(-\half\int (\dd g)^4\;|\sigma_0(g_I)|^2\right)\,.
\eena
As we will see in the following, while this type of perturbation does not allow us to go away from spatial homogeneity, it already gives several conceptual insights strengthening the link between the effective dynamics of condensate states and usual quantum cosmology.

Using 
\ben
\hphi(g_I)|\tau,\sigma_0\rangle = \frac{\mathcal{N}(\tau,\sigma_0)}{\mathcal{N}(\sigma_0)}\,\tau(g_I)|\sigma_0\rangle + \sigma_0(g_I)|\tau,\sigma_0\rangle
\een
we find that
\bena
&&\int (\dd g)^4\;\langle\tau,\sigma_0|\hphid(g_I)\mathcal{K}\hphi(g_I)|\tau,\sigma_0\rangle
\\& = & \frac{\mathcal{N}(\tau,\sigma_0)^2}{\mathcal{N}(\sigma_0)^2}\int (\dd g)^4 \;\bar\tau(g_I)\mathcal{K}\tau(g_I)\nonumber
\\&& + \frac{\mathcal{N}(\tau,\sigma_0)}{\mathcal{N}(\sigma_0)}\left(\langle\sigma_0|\tau,\sigma_0\rangle\int (\dd g)^4\;\bar\tau(g_I)\mathcal{K}\sigma_0(g_I)+{\rm c.c.}\right)\nonumber
\\&&+\int (\dd g)^4 \;\bar\sigma_0(g_I)\mathcal{K}\sigma_0(g_I)\,.\nonumber
\eena
The overlap between the unperturbed and perturbed condensate states is
\bena
\langle\sigma_0|\tau,\sigma_0\rangle &=& \frac{\mathcal{N}(\tau,\sigma_0)}{\mathcal{N}(\sigma_0)}\int (\dd g)^4\; \tau(g_I)\,\langle\sigma_0|\hphid(g_I)|\sigma_0\rangle\nonumber
\\& = &\frac{\mathcal{N}(\tau,\sigma_0)}{\mathcal{N}(\sigma_0)}\int (\dd g)^4\; \tau(g_I)\bar\sigma_0(g_I)
\eena
and we finally obtain
\bena
&&\int (\dd g)^4\;\langle\tau,\sigma_0|\hphid(g_I)\mathcal{K}\hphi(g_I)|\tau,\sigma_0\rangle\nonumber
\\& = &\frac{\mathcal{N}(\tau,\sigma_0)^2}{\mathcal{N}(\sigma_0)^2}\left(\int (\dd g)^4 \;\bar\tau(g_I)\mathcal{K}\tau(g_I)\right.\nonumber
\\&& + \int (\dd g)^4\;\tau(g_I)\bar\sigma_0(g_I)\int (\dd h)^4\;\bar\tau(h_I)\mathcal{K}\sigma_0(h_I)\nonumber
\\&&\left.+\int (\dd g)^4\;\bar\tau(g_I)\sigma_0(g_I)\int (\dd h)^4\;\bar\sigma_0(h_I)\mathcal{K}\tau(h_I)\right)\nonumber
\\&&+\int (\dd g)^4 \;\bar\sigma_0(g_I)\mathcal{K}\sigma_0(g_I)
\label{condition}
\eena
where we are assuming that $\mathcal{K}$ is self-adjoint. Note that
\ben
\frac{\mathcal{N}(\tau,\sigma_0)^2}{\mathcal{N}(\sigma_0)^2} = \frac{1}{\left|\int (\dd g)^4\;\bar\tau(g_I)\sigma_0(g_I)\right|^2+\int (\dd g)^4 \;|\tau(g_I)|^2}\,.
\een
Setting $\mathcal{K}\equiv 1$ we find the average particle number
\bena
N&=&\int (\dd g)^4\;\langle\tau,\sigma_0|\hphid(g_I)\hphi(g_I)|\tau,\sigma_0\rangle\nonumber
\\& = &\int (\dd g)^4 \;|\sigma_0(g_I)|^2 + 1\nonumber
\\&& + \frac{\left|\int (\dd g)^4\;\bar\tau(g_I)\sigma_0(g_I)\right|^2}{\left|\int (\dd g)^4\;\bar\tau(g_I)\sigma_0(g_I)\right|^2+\int (\dd g)^4 \;|\tau(g_I)|^2}\,.
\eena
The perturbation by $\tau$ increases the average particle number by between 1 and 2, depending on the overlap between the wave functions $\tau$ and $\sigma_0$ in minisuperspace. It is indeed a very small perturbation if, as we are assuming throughout, $N_0=\int (\dd g)^4 \;|\sigma_0(g_I)|^2\gg 1$.

If we now require, as before, that the expectation value of the GFT ``kinetic energy'' vanishes,
\ben
\int (\dd g)^4\;\langle\tau,\sigma_0|\hphid(g_I)\mathcal{K}\hphi(g_I)|\tau,\sigma_0\rangle = 0\,,
\een
we find four different terms from Eq.~(\ref{condition}). Comparing the terms inside the brackets, we see that the first should be negligible for a large enough number of particles in the background (so that $|\sigma_0(g_I)|\gg 1$). We then obtain
\begin{widetext}
\ben
\int(\dd g)^4\;\bar\sigma_0(g_I)\int (\dd h)^4\left(\frac{\tau(g_I)\bar\tau(h_I)\mathcal{K}_h+(\mathcal{K}\tau)(g_I)\bar\tau(h_I)}{\left|\int (\dd g')^4\;\bar\tau(g'_I)\sigma_0(g'_I)\right|^2+\int (\dd g')^4 \;|\tau(g'_I)|^2}+\delta^4(g_I^{-1}h_I)\mathcal{K}_h\right)\sigma_0(h_I)\approx 0\,.
\label{backrham}
\een
\end{widetext}
By including the perturbation by $\tau$, the resulting effective dynamics for the background wave function $\sigma_0$ is modified by nonlocal and highly nonlinear terms that depend on the perturbation wave function, leading to a nonlocal effective ``Hamiltonian constraint.'' These nonlocalities on minisuperspace encode the backreaction of inhomogeneities on the background. Their effect is very small, as expected; due to the extra factor of $\sigma_0^{2}$ in the denominator, it is generically of order $1/N_0$ relative to the unperturbed background evolution (though it can be large for very special states, e.g. states with no overlap of $\tau$ and $\sigma_0$ in minisuperspace).

\section{Effective quantum cosmological dynamics}

In a situation where there is a very small perturbation to the dynamics of a background wave function, one can study the system by solving for background and perturbation separately. Hence let us imagine solving Eq.~(\ref{backrham}) for the background $\sigma_0$ ignoring the nonlinearities. Equation (\ref{condition}) can then be read as specifying an effective dynamics for the perturbation $\tau$,
\bena
\int(\dd g)^4\;\bar\tau(g_I)\int (\dd h)^4\left(\sigma_0(g_I)\bar\sigma_0(h_I)\mathcal{K}_h\right.&&\nonumber
\\\left.+(\mathcal{K}\sigma_0)(g_I)\bar\sigma_0(h_I)+\delta^4(g_I^{-1}h_I)\mathcal{K}_h\right)\tau(h_I)&\approx& 0\,.
\label{taudyn}
\eena
Again, the last (local) piece is completely negligible for a large enough number of background quanta $N_0$. For general $\sigma_0$, the dynamics for $\tau$ appears highly nonlocal in minisuperspace. Note however that we have made no assumptions about the form of $\sigma_0$, so that it does in general not correspond to any background geometry on which fluctuations could propagate. Equation (\ref{taudyn}) describes the interaction between a quantum condensate and a quantum perturbation of it, and does not in general admit a semiclassical picture. If we assume a $\sigma_0$ that is sharply peaked around some $g_I^0$ [or rather an equivalence class $\{g_I^0 h\}$ under (\ref{gaugeinv})], these nonlocalities will be strongly suppressed. (Nonlocalities involving configurations that are related by a gauge transformation are not physical; one can work on a smaller configuration space of gauge-invariant geometric data as outlined in Ref.~\cite{steffegfc}.)

If we discard the local term quadratic in $\tau$, there is another interesting observation. Defining a new wave function $\Psi(g_I,h_I)$ describing both the background and the perturbation by $\Psi(g_I,h_I):=\sigma_0(g_I)\tau(h_I)$, Eq.~(\ref{taudyn}) can be recast as
\ben
\int (\dd g)^4 (\dd h)^4\;\overline{\Psi}(g_I,h_I)\mathcal{P}(\mathcal{K}_h+\mathcal{K}_g)\Psi(g_I,h_I)\approx 0
\label{effham}
\een
where $\mathcal{P}$ is a permutation $\mathcal{P}f(g_I,h_I):=f(h_I,g_I)$. In this form, the dynamics can be interpreted as an effective Hamiltonian on an enlarged minisuperspace, spanned by $(g_I,h_I)$, of the possible geometric configurations of background and perturbation combined. The resulting dynamics is then evidently reminiscent of the Born--Oppenheimer approximation for small perturbations of an FLRW universe in usual quantum cosmology. For full details of the reasoning behind this approximation, see e.g. the textbook \cite{kieferbook}; here we are summarizing the most important points. Given a wave function on a ``superspace'' encoding both the FLRW background and perturbations over it, and dynamics given by the Hamiltonian constraint of general relativity expanded up to second order in the perturbations, one regards the background variables as ``heavy'' and the perturbations as ``light'' in the Born--Oppenheimer sense. The wave function is taken to be (\ref{prodform}) where the dependence of the fluctuation wave functions $\Psi^{(n)}$ on the background variables $(a,\phi)$ is adiabatic, i.e. derivatives of $\Psi^{(n)}$ with respect to $a$ and $\phi$ are small,
\ben
\left|\frac{\partial\Psi_0}{\Psi_0}\right| \gg \left|\frac{\partial\Psi^{(n)}}{\Psi^{(n)}}\right|
\label{ass}
\een
for $\partial=\partial/\partial a$ or $\partial=\partial/\partial \phi$. One then splits the equation $\mathcal{H}\Psi=0$ into a Wheeler--DeWitt equation for the background $\Psi_0$ and equations for each of the $\Psi^{(n)}$. By using (\ref{ass}) and the assumption that $\Psi_0$ is of WKB form, an effective ``Schr\"odinger equation'' for $\Psi^{(n)}$ emerges, with an approximate ``WKB time'' defined through $\Psi_0$.

In our setting no metric superspace or Hamiltonian on it exists {\em a priori}, but both are reconstructed from the quantum dynamics of the GFT condensate and the perturbation over it. A Born--Oppenheimer approximation would correspond to a wave function
\ben
\Psi_{{\rm B.O.}}(g_I,h_I):=\sigma_0(g_I)\tau(h_I;g_I)
\een
for which derivatives of $\tau$ with respect to $g_I$ are assumed to be small, and a split of the full dynamics into separate equations for the background and perturbations. Here, we recover precisely this structure in the dynamics of the background wave function $\sigma_0$ and the product wave function $\Psi$ that includes perturbations, in the limit where one neglects derivatives of $\tau$ with $g_I$ so that $\tau$ is only a function of $h_I$. The analogue of the Born--Oppenheimer criterion of a separation into heavy and light degrees of freedom is the condition $N_0\gg 1$ that implies that the perturbation by $\tau$ really is a small perturbation of the homogeneous condensate. This condition also implies that the reconstructed (approximate) metric describing the condensate and its perturbation is only minimally affected by the perturbation: as discussed in Ref.~\cite{atomnumber}, such a metric is reconstructed from geometric quantities that are {\em extensive} observables of the condensate. Thus if $1/N_0\ll 1$ we have a small perturbation $\delta g/g \ll 1$. Note that conditions like $N_0\gg 1$ can be verified for any given condensate state such as Eq.~(\ref{pertcond}), in contrast to conventional quantum cosmology where the consistency of a Born--Oppenheimer approximation must be assumed.

In order to complete the interpretation of Eq.~(\ref{effham}) as giving the dynamics for an effective Born--Oppenheimer quantum cosmology wave function, one should show that the effective Hamiltonian appearing in Eq.~(\ref{effham}) can be interpreted, at least in an appropriate semiclassical limit, as a gravitational Hamiltonian constraint for a homogeneous background metric with a small perturbation. This is what we will investigate next.

Equation (\ref{effham}) can be interpreted as requiring the expectation value of the operator $\mathcal{P}(\mathcal{K}_h+\mathcal{K}_g)$ to (approximately) vanish in the state specified by the first quantized quantum cosmology wave function $\Psi$. One needs to interpret this operator in terms of cosmological variables. 

The operator $\mathcal{P}$ corresponds to a permutation of the background and perturbation variables on minisuperspace, and its interpretation in terms of quantum cosmology is not totally clear. Its appearance may be related to the corresponding classical symmetry that, for a spatially homogeneous perturbation (see below), the splitting between background and perturbation is completely arbitrary. If one imposes a symmetry under $\mathcal{P}$ on the combined wave function $\Psi$, $\mathcal{P}$ would disappear from the expectation value (\ref{effham}).

For concreteness, we can set $\mathcal{K}=\sum_I \Delta_{g_I}+m^2$ as was done in the previous work of Refs.~\cite{gfcpapers,steffegfc,atomnumber}. This type of kinetic term appears naturally in the renormalization of GFT models \cite{GFTrenorm}. If, as in Ref.~\cite{steffegfc}, we restrict to isotropic configurations for which the wave function $\Psi$ only depends on one isotropic combination of group elements $p_0$ for the background and another one $q$ for the perturbation, we have
\bena
&&(\mathcal{K}_h+\mathcal{K}_g)\Psi(g_I,h_I)\nonumber
\\& = &\left(2p_0(1-p_0)\frac{\partial^2}{\partial p_0^2}+(3-4p_0)\frac{\partial}{\partial p_0}+2q(1-q)\frac{\partial^2}{\partial q^2}\right.\nonumber
\\&&\left.+(3-4q)\frac{\partial}{\partial q}+\frac{m^2}{2}\right)\Psi(p_0,q)
\label{waveeq}
\eena
as the explicit expression of the differential operator $(\mathcal{K}_h+\mathcal{K}_g)$ in terms of isotropic minisuperspace coordinates $(p_0,q)$. Note that the ``effective Wheeler--DeWitt equation'' $(\mathcal{K}_h+\mathcal{K}_g)\Psi=0$ is a sufficient but not necessary condition for Eq.~(\ref{effham}) to hold; Eq.~(\ref{effham}) would correspond to a weakly imposed Hamiltonian constraint in the quantum cosmology picture. As outlined in Sec.~\ref{condenintro}, Eq.~(\ref{effham}) is just one out of an infinite number of expectation values coming from the Schwinger--Dyson equations of the GFT, and finding an exact physical state of the GFT would (in theory) amount to solving many more consistency relations than one Wheeler--DeWitt equation in quantum cosmology. Solutions to Eq.~(\ref{effham}) can provide a first approximation to such an exact physical state.

As a second-order differential operator, Eq.~(\ref{waveeq}) defines a metric on the minisuperspace parametrized by $(p_0,q)$; one can absorb the first derivatives in a coordinate redefinition $P_0=P_0(p_0)$, $Q=Q(q)$ and then read off the (inverse) metric as the coefficients in front of the second derivatives. Explicitly, choosing
\ben
P_0=\sqrt{\frac{1-p_0}{p_0}}\,,\quad Q=\sqrt{\frac{1-q}{q}}\,,
\label{coortraf}
\een
Eq.~(\ref{waveeq}) becomes
\bena
&&(\mathcal{K}_h+\mathcal{K}_g)\Psi(g_I,h_I)
\label{wave2}
\\& = &\left((1+P_0^2)\frac{\partial^2}{\partial P_0^2}+(1+Q^2)\frac{\partial^2}{\partial Q^2}+m^2\right)\Psi(P_0,Q)\,.\nonumber
\eena
This defines a minisuperspace metric which is diagonal and nondegenerate everywhere except at the boundary points $P_0=\infty$ or $Q=\infty$ (note that while $p_0$ and $q$ are in the unit interval $[0,1]$, the range of $P_0$ and $Q$ is $[0,\infty]$).

In this canonical form, Eq.~(\ref{wave2}) can be used to introduce a notion of ``WKB time'' on minisuperspace as for the Born--Oppenheimer approximation in usual quantum cosmology. This time is well defined in a regime in which the wave function $\sigma_0$ is taken to be of WKB form, so that
\ben
\Psi(P_0,Q)=e^{\im S_0(P_0)/\hbar {\rm G}}\chi(P_0,Q)
\een
where $S_0$ oscillates very rapidly compared to $\chi$ (we have absorbed the slowly varying absolute value of $\sigma_0$ into $\chi$). One can then approximate
\ben
(1+P_0^2)\frac{\partial^2\Psi}{\partial P_0^2}\approx \frac{\im}{\hbar{\rm G}}\,(1+P_0^2)\frac{\partial S_0}{\partial P_0}\,\frac{\partial \Psi}{\partial P_0}\,;
\een
identifying $\partial S_0/\partial P_0$ with the (WKB) momentum conjugate to $P_0$, by using Hamilton's equations for the Hamiltonian defined by Eq.~(\ref{wave2}) one then notices that the combination $(1+P_0^2)\,\partial S_0/\partial P_0$ is equal to $-\half$ times the time derivative of $P_0$. This defines a time $t$ by
\ben
(1+P_0^2)\frac{\partial^2\Psi}{\partial P_0^2}\approx -\frac{\im}{2\hbar{\rm G}}\,\frac{{\rm d}P_0}{{\rm d}t}\frac{\partial \Psi}{\partial P_0}\approx -\frac{\im}{2\hbar{\rm G}}\,\frac{\partial \Psi}{\partial t}
\een
again using the WKB assumption that derivatives with respect to $P_0$ dominate the gradient of $\Psi$. All this here formally goes through as in standard quantum cosmology, but relies on having a background wave function $\sigma_0$ of WKB form; as discussed above and again in the following, it remains unclear whether this assumption is satisfied for physically interesting GFT condensate states.

In order to interpret the coordinates $p_0$ and $q$ on the group in terms of a gravitational connection, the scaling of an ``averaged holonomy,'' to be associated to such a connection, with the number of GFT quanta must be taken into account \cite{atomnumber}. One can then identify $p_0$ and $q$ with parallel transports of a gravitational connection,
\ben
p_0 \propto \sin^2 (\nu\,N^{-1/3}\,\omega_p)\,,\quad q \propto \sin^2 (\nu\,N^{-1/3}\,\omega_q)
\een
where $\nu$ is a free parameter and $N$ is the average number of quanta in the condensate \cite{atomnumber}. Trivial parallel transport $p_0=0$ corresponds to a flat connection whereas $p_0\sim O(1)$ means large curvature on the (perhaps Planckian) scale set by the discrete quanta. Equation (\ref{coortraf}) then means
\ben
P_0 \propto \cot (\nu\,N^{-1/3}\,\omega_p)\,,\quad Q \propto \cot (\nu\,N^{-1/3}\,\omega_q)
\een
so that a flat connection corresponds to $P_0=\infty$ or $Q=\infty$, which seems less convenient. We therefore use the variables $(p_0,q)$ and Eq.~(\ref{waveeq}) in the following.

The detailed interpretation of Eq.~(\ref{waveeq}) as an effective Friedmann equation depends not only on the interpretation of such coordinates on minisuperspace but also on how $N$ scales with other cosmological variables such as the scale factor. Furthermore, while one may employ a semiclassical WKB-type approximation in which the highest derivatives dominate Eq.~(\ref{waveeq}), the interpretation of this approximation is not clear as it assumes that the {\em average} area per quantum of geometry is large compared to the elementary (presumably Planckian) area scale in the theory. See Refs.~\cite{steffegfc,atomnumber,gianluca} for discussions of the WKB approximation in this context. 

The effective Hamiltonian for the wave function $\Psi$ given by Eq.~(\ref{effham}) is a sum of decoupled kinetic terms for background and perturbations. To be more concrete, we have to interpret Eq.~(\ref{effham}) in terms of cosmological dynamics, i.e. an effective Friedmann equation. As said before, this interpretation depends rather crucially on the behavior of the atomic number $N$ relative to other cosmological variables. Interpreting Eq.~(\ref{effham}) in terms of expectation values and assuming a relation $N=N(a)$, various possibilities for the effective cosmological dynamics were discussed in Ref.~\cite{atomnumber}. It is then clear that, for any effective Friedmann equation for the background variables derived in this way, Eq.~(\ref{effham}) simply gives the sum of two such Friedmann equations for the background and perturbation variables.

One possible, rather crude, derivation of this sort would be to apply a WKB approximation to Eq.~(\ref{waveeq}) taking only the highest derivatives into account. In this limit, requiring a zero expectation value for $\mathcal{K}_h+\mathcal{K}_g$ means that the coefficients in front of the highest derivatives have to be tuned to zero in terms of the WKB variables,
\ben
2p_0(1-p_0)\approx 0\,,\quad 2q(1-q) \approx 0\,.
\label{approx}
\een
Out of the two solutions for each equation, only one is viable in the geometric interpretation of GFT condensates. $p_0$ and $q$ represent the spatial curvature measured on the scale of an individual tetrahedron, which must be small for the condensate to approximate a continuum geometry. This enforces the only allowed solution
\ben
p_0\approx 0\,,\quad q\approx 0\,.
\label{zerocons}
\een
$p_0\approx 1$ or $q\approx 1$ would correspond to a geometry with large curvature on presumably Planckian scales. In this approximation, one would conclude from Eq.~(\ref{zerocons}) that the connection given by $\omega_p$ and $\omega_q$ has to be flat and the semiclassical solution is Minkowski spacetime, suggesting that the classical limit of Eq.~(\ref{effham}) is compatible with the dynamics of classical vacuum GR (where this would be the only solution). However, we should mention again that the viability of such WKB approximations in the study of GFT condensate states has been questioned by the analysis of Ref.~\cite{steffegfc}, as it means assuming large microscopic average areas, as discussed in Ref.~\cite{atomnumber}. A more detailed study is needed to see whether there exist well-behaved and physically relevant states solving Eq.~(\ref{effham}) that are also peaked on the classical values (\ref{zerocons}). In Ref.~\cite{steffegfc} it was shown, for wave functions only depending on one geometric variable $p_0$, that depending on the value of $m^2$ there may or may not exist such solutions, and solutions that do exist generally do not display rapid oscillation as assumed in the WKB limit. The argument presented here should therefore be seen as very tentative.

\section{Discussion}

We have given a tentative argument suggesting that a classical limit of the effective quantum cosmological dynamics, for a perturbed GFT condensate of the form (\ref{pertcond}), could indeed reproduce the expectations from vacuum GR if the perturbation is interpreted as spatially homogeneous. As discussed in Ref.~\cite{atomnumber}, a more careful analysis that does not require the WKB approximation will generally give corrections to the vacuum Friedmann equation which depend on assumptions about the scaling relation $N(a)$. More work is then needed to interpret the specific form of the dynamics given by Eq.~(\ref{effham}) for different choices of GFT dynamics and condensate states. 

The main point of this paper is however more general than this. We have given a consistent interpretation of the simplest perturbation of an exactly homogeneous condensate as a spatially homogeneous metric perturbation. The quantum dynamics of such a state is controlled by the effective Hamiltonian (\ref{effham}) for a product wave function, and an effective Born--Oppenheimer approximation emerges from the smallness of the average particle number of the perturbation with respect to the number of quanta in the background $N_0$. The same smallness also guarantees that the reconstructed geometry is indeed a very small perturbation of the geometry determined by the background condensate state. 

The interpretation of one or two atoms perturbing the condensate as a spatially homogeneous perturbation is perfectly consistent with the geometric interpretation of general GFT Fock states introduced in Refs.~\cite{gfcpapers}. Imagine reconstructing a metric geometry by embedding the background condensate into a manifold. Then, add a perturbation, which will be embedded somewhere in the manifold, so that one should reconstruct a spatial geometry that is spatially homogeneous everywhere except one small patch in which it looks different. However, one also has to take into account quantum-mechanical indistinguishability of the bosonic GFT quanta, meaning that the reconstructed geometry is in fact a superposition of all permutations of the chosen embedding. But this implies that the perturbation cannot be localized in the embedding space; it can be found at any of the embedding patches with equal probability. Hence, if any type of semiclassical description is viable, it should be that of a spatially homogeneous metric perturbation.

Of course, classically this perturbation can simply be absorbed into the background, and so at first glance does not seem to add anything of interest to the unperturbed condensate. However, we now have a different quantum system with twice as many degrees of freedom; in the isotropic case, $a$ and $L$ are independent variables, and can fluctuate independently. Including anisotropies in the perturbation $\tau$, with a background condensate assumed to be isotropic, allows for a systematic perturbative treatment of anisotropies. More generally and more importantly, studying this very simplest type of perturbation already shows how an effective formalism for quantum cosmology of a homogeneous background with perturbations can emerge from suitable states in a fundamental quantum gravity theory given by GFT. Background and perturbations can be distinguished by computing the average particle number, as we have shown. The very existence of such a discrete quantum observable is a genuine quantum gravity effect, whose possible cosmological consequences have been already anticipated in Ref.~\cite{atomnumber} (a dependence of quantum gravity corrections in LQC on this observable has also been observed in Ref.~\cite{emafrancN}). Here, it allows us to set up an effective Born--Oppenheimer approximation in which it is justified to first solve the equations for the background, ignoring nonlocal and nonlinear terms coming from backreaction of the perturbations, and then to define a wave function of product type whose dynamics is governed by an effective Hamiltonian on an enlarged minisuperspace including the degrees of freedom of both background and perturbations. While indications that the dynamics has the correct semiclassical limit corresponding to a vacuum universe in classical general relativity will require further analysis to become conclusive, the technical and conceptual insights gained in our analysis should be helpful in future work towards understanding the major open issue of including inhomogeneous perturbations in quantum gravity condensates.

It is clear from the previous discussion that, in developing such a formalism for inhomogeneities, one will need to consider more complicated states than the very simple ones given by Eq.~(\ref{pertcond}). Inhomogeneous perturbations must be localizable with respect to the background, while in our setting both the background and the perturbation describe a fluid made up of a large number of identical, uncorrelated excitations. This calls for an extension of our approach to one using states that contain correlations between different quanta. In the discrete geometry language of GFT, such correlations encode topological information; instead of only considering disconnected ``atoms'' or small ``molecules'' as we have done here, one could consider states describing large connected structures of many such GFT quanta. A concrete method for constructing such states was put forward in Ref.~\cite{newcondensates}. The interpretation of the states of Ref.~\cite{newcondensates} requires no embedding into an (arbitrary) manifold; such states form a macroscopic (simplicial) manifold of their own, with topology determined by combinatorial data contained in the state. We can envisage using such a condensate state, say with topology of a 3-sphere, to define an intrinsic notion of spherical harmonics, only making reference to the background condensate, with respect to which (a more complicated notion of) inhomogeneous fluctuations can then be defined. The limitations of the simple type of perturbations of GFT condensates considered in this paper then are mainly a consequence of the simple ansatz (\ref{pertcond}) and not a feature of the general program of extracting quantum cosmology from condensate states in GFT.

\section*{Acknowledgments}

I would like to thank E. Wilson-Ewing and a referee for helpful comments. The research leading to these results has received funding from the People Programme (Marie Curie Actions) of the European Union's Seventh Framework Programme (FP7/2007-2013) under REA grant agreement n$^{\rm o}$ 622339. Research at Perimeter Institute is supported by the Government of Canada through Industry Canada and by the Province of Ontario through the Ministry of Research and Innovation.


\end{document}